\documentclass[twocolumn]{aa}
\usepackage{txfonts}
\usepackage{natbib}
\usepackage{balance}
\usepackage{graphicx}
\usepackage[a4paper]{hyperref}
\begin{document}
\def\kms{$\mathrm {km s}^{-1}$}

   \title{A search for Interstellar absorptions towards Sgr dSph}

   \subtitle{}

   \author{S. Monai
          \inst{1}
          \and
          P. Bonifacio\inst{1} 
\and
L. \,Sbordone  \inst{2,3}
\thanks{Based on observations obtained with
	  UVES at VLT Kueyen 8.2m telescope in program 67.B-0147}
          }

   \offprints{S. Monai}

   \institute{Istituto Nazionale di Astrofisica -  Osservatorio Astronomico di
   Trieste, Via Tiepolo 11 34131 Trieste -Italy
\and
European Southern Observatory, Casilla 19001, Santiago, Chile
\and 
Universit\`a di Roma 2 "Tor Vergata", via della Ricerca Scientifica, Rome, Italy}

   \date{}

   \abstract{We searched for 
   Na I D1,D2 interstellar absorption lines
   towards 12 giant stars 
  of the Sgr dwarf spheroidal and 3 stars of 
the associated  globular cluster Terzan 7, observed at high
   resolution with the UVES spectrograph 
at the 8.2 m Kueyen VLT telescope. These stars are
   not an ideal background source for the  analysis
of interstellar (IS) absorption lines since  they have been observed 
   in the framework of a stellar abundances study. However they are 
   sufficiently luminous to allow a decent 
   S/N ratio at a resolution of about 43000,
   i.e. ~7 \kms, which allows an exploratory study.
   We detected two distinct groups of IS absorptions, 
a  ``local'' group,  with
radial velocities ranging 
   from -43 \kms ~ to 50 \kms, and a high velocity group 
with radial velocities ranging from 150 km/s to
   165 km/s. Likely the high velocity components are due to 
   gas falling on the Sgr dSph
   since have been  observed  
only in 2 stars which are separated by only
   1\farcm{5} in the sky. We argue that the observations suggest a
   stripping of gas due to the passage of Sgr through the Galactic disc. 
   
   }

   \maketitle
%
%________________________________________________________________

\begin{table*}
\caption{Basic data on the stars used as background sources}
\label{basic}
\begin{center}
\begin{tabular}{lrrrrrrrrrr}
\hline
\\
Star$^a$ & \multispan3{\hfill$\alpha(2000)^b$\hfill} &  \multispan3{\hfill$\delta(2000)^b$\hfill}&
\multispan2{\hfill$V~~~ (V-I)_0^c$} &
$v_{rad}$ & $\sigma_v$  \\
   \multispan9{~}& \kms & \kms\\
\\
\hline
\\
432      &18& 53& 50.75&  $-30$& 27& 27.3&  17.55 & 0.965 & 160.8 & 0.3 \\%124bcd
628      &18& 53& 47.91&  $-30$& 26& 14.5&  18.00 & 0.928 & 145.7 & 0.6\\ %141abc
635      &18& 53& 51.05&  $-30$& 26& 48.3&  18.01 & 0.954 & 125.2 & 0.2\\  %122abc
656      &18& 53& 45.71&  $-30$& 25& 57.3&  18.04 & 0.882 & 135.0 & 0.3\\  %159abd
709      &18& 53& 38.73&  $-30$& 29& 28.5&  18.09 & 0.917 & 125.2 & 0.5\\  %201abcd
716      &18& 53& 52.97&  $-30$& 27& 12.8&  18.10 & 0.902 & 136.0 & 0.4\\  %105abc
717      &18& 53& 48.05&  $-30$& 29& 38.1&  18.10 & 0.872 & 142.0 & 0.4\\  %142abc
772$^d$  &18& 53& 48.13&  $-30$& 32&  0.8&  18.15 & 0.947 & 143.8 & 0.8\\   %143
867      &18& 53& 53.02&  $-30$& 27& 29.2&  18.30 & 0.933 & 154.3 & 0.2\\   %104abc
879$^e$  &18& 53& 48.59&  $-30$& 30& 48.7&  18.33 & 0.965 & 133.8 & 0.8\\   %139
894      &18& 53& 36.84&  $-30$& 29& 54.1&  18.34 & 0.940 & 129.5 & 0.8\\   %210abcd
927      &18& 53& 51.69&  $-30$& 26& 50.7&  18.39 & 0.937 & 144.0 & 0.3\\   %115abcd
1282$^f$ &19& 17& 39.36&  $-34$& 39&  6.4&  16.08 & 1.324 & 158.2 & 1.0\\      
1515$^f$ &19& 17& 38.16&  $-34$& 29& 16.8&  16.76 & 1.156 & 157.8 & 1.0\\      
1272$^f$ &19& 17& 37.11&  $-34$& 39& 11.8&  16.62 & 1.188 & 158.9 & 1.0\\      
\\
\hline
\end{tabular}
\\
\end{center}
\hbox{$^a$ 
 Star numbers for Sgr stars are from \citet{Marconi} field 1, available
through CDS\hfill}
\hbox{ at {\tt cdsarc.u-strasbg.fr/pub/cats/J/A+A/330/453/sagit1.dat}\hfill}
\hbox{$^b$  accurate to 0\farcs{3} (Ferraro \& Monaco 2002, priv. comm. 
for Sgr and Zaggia 2004 priv.comm. for Terzan 7)\hfill}
\hbox{$^c$ The adopted reddening is $E(V-I)=0.22$ }
\hbox{$^d$ this is star  [BHM2000] 143  of \citet{B00}\hfill}
\hbox{$^e$ this is star  [BHM2000] 139  of \citet{B00}\hfill}
\hbox{$^f$ Star numbers for Terzan 7  stars are from the catalogue 
of \citet{Buonanno} available through CDS\hfill}
\hbox{ at {\tt cdsarc.u-strasbg.fr/pub/cats/J/AJ/109/663/table2.dat}. \hfill}

\end{table*}

\section{Introduction}
The Sagittarius dwarf spheroidal galaxy (Sgr dSph) 
is one of the nearest satellites of the Milky
Way. It offers an interesting opportunity for investigating the structure of 
the interstellar medium (ISM) in the Galactic Halo and, in principle,
along all the line of sight up to the dwarf galaxy.
The UVES spectra  of 12 giant stars
in Sgr and 3 giant stars in the associated  
globular cluster Terzan 7, which have been
recently used to investigate the chemical composition of the
Sgr dSph \citep{B00,Bonifacio,sbordone} give us the
opportunity to perform the first study of this kind.
The Sgr giants are, of course, not ideal background sources
for the study of the ISM, since the spectrum is complicated
by the many stellar photospheric absorption lines. 
The only suitable ISM lines for which such a study is possible,
with the above data,
are the NaI D doublet lines.  The observed stars
are K giants and display
strong photospheric absorptions in these lines.
However the large radial velocity of Sgr ($\sim 140$ \kms) guarantees
that at least all the low radial velocity Galactic
absorptions are not affected by the corresponding stellar absorptions.
Luckily, the spectral region adjacent to the Na I D lines is relatively
free from absorptions in the stellar spectrum, these
two conditions allow to use these spectra for an exploratory
study of the ISM towards Sgr.
We further note that the hot stars in Sgr (the so called ``Blue Plume''
and the HB stars), which would
be more desirable background sources
for the study of the ISM,  
are considerably fainter,
% with  magnitudes in the range
%V=19--20, 
which implies that their use as background sources
requires a major investment in telescope time.
It is thus of paramount importance to make the best
possible use of the available spectra.

\begin{table}[tbh]
   \caption{Column densities, b factors and velocities, for our sample. 
   * means that $b$ and column density
were assumed and not fitted.} 
\label{tabstars}
\centering
\begin{tabular}{lllr}
\hline
{ Star} & { logN} &  { $b$} &  v   \\ 
           &            &   \kms     & \kms \\ 
\hline
 867  & 13.42 &  0.50 &$  -17 $ \\
      & 14.63 &  1.57 &$  -2  $ \\
 716  & 11.58 &  6.05 &$   11 $ \\
      & 11.48 &  7.33 &$   34 $ \\
 927  & 12.43 &  6.81 &$  -15 $ \\
      & 12.42 &  3.70 &$  -7  $ \\
      & 11.94 &  1.45 &$   9  $ \\
 635  & 11.80* & 0.60 &$ 150  $\\
      & 11.50* & 0.50 &$ 165  $\\
 432  & 12.47* & 0.93 &$ -16  $\\
      & 13.84* & 1.30 &$  -5  $\\
      & 13.00* & 1.10 &$   7  $\\
 879  & 12.40* & 5.30 &$  -1  $ \\
      & 12.60* & 1.40 &$  18  $\\
      & 12.00* & 0.20 &$  40  $\\
      & 12.00* & 0.20 &$  50  $\\
 628  & 13.90* & 0.20 &$  -4  $\\
      & 12.40* & 0.68 &$  10  $\\
      & 11.50* & 1.70 &$  28  $\\
      & 11.60* & 0.30 &$  36  $\\
      & 11.40* & 0.30 &$  44  $ \\
 717  & 12.45* & 0.28 &$ -15  $ \\
      & 12.60* & 0.33 &$  -5  $   \\
      & 12.45* & 0.58 &$  14  $  \\   
 772  & 13.20* & 2.08 &$  -2  $   \\
      & 12.46* & 3.43 &$  17  $   \\
      & 11.71* & 2.11 &$  26  $   \\
 656  & 12.10* & 1.70 &$ -17  $    \\
      & 13.40* & 1.90 &$  -5  $    \\
      & 11.50* & 0.67 &$ 160  $ \\
 709  & 12.64 &  0.94 &$ -16  $       \\
      & 13.15 &  2.41 &$  -2  $      \\
      & 12.18 &  3.63 &$  15  $    \\
      & 14.53 &  0.00 &$ 36   $  \\
 894  & 11.80* &  1.40 &$ -16 $  \\
      & 12.40* &  2.75 &$  -5 $   \\
      & 11.80* &  1.70 &$   8 $   \\
      & 12.60 &  0.40  &$  20 $   \\
1282  & 11.87 &  2.82 & $ -42  $          \\
      & 13.72 &  2.29 & $  1   $        \\
1515  & 11.60 &  2.04 & $ -40  $         \\
      & 12.70* & 3.90 & $  0   $         \\
1272  & 12.38 &  2.33 & $ -43  $          \\
      & 13.79 &  2.34 & $  0   $          \\
\hline
\end{tabular} 
\end{table}

In general dSph galaxies are considered to be
gas poor, as opposed to  dwarf irregulars.
With regard to the gas content of the Sgr dSph, 
pointed HI observations of its
central region indicate that it does not contain a significant amount of 
neutral gas
\citep{Koribalski}. The search for HI was continued by 
\citet{Burton}, who  found no detectable
emission and provided an upper limit to the 
HI mass $\rm M(HI) <7000 M_\odot$ 
over 18 deg$^2$, between
$b = -13^{\circ}$ and $-18^{\circ}.5$. 
This result is somewhat surprising
since  the chemical abundances \citep{B00,Bonifacio} and the
colours of the Sgr stars show that it comprises
a metal-rich young population 
implying that the
Sgr dSph was forming stars within the last Gyr. 
\citet{Bonifacio} suggested that the passages
of Sgr through the Galactic disc trigger star formation episodes,
however this implies that there is gas available in Sgr or
that it can be stripped from the disc.
The ``Blue Plume'' stars are members of Sgr \citep{zaggia}
and there is little question that they are the Main Sequence
of the young population. Still the question lingers: where
has the gas associated to this recent star formation gone?
A recent investigation of 
\citet{Putman} has shown that gas is associated 
with the stream of stellar
debris found to extend, along the Sgr dSph orbit, 
for over $150^{\circ}$ across the 
south Galactic hemisphere.
It thus appears that  the tidal interaction
of Sgr with the Milky Way results in the stripping of gas. 
\par 
Having this in mind,
our attention has been focused on the search 
of ISM components at radial velocities close to 
those of the Sgr dSph stars.  
This is certainly a necessary condition for gas belonging
to Sgr, although radial velocity by itself
is not sufficient to prove that the gas is actually  at that distance.
 
%__________________________________________________________________

\section{Observations and Analysis}

Details on the observations 
are reported in  \citet{B00}, \citet{Bonifacio} and \citet{sbordone}. 
For the convenience of the reader we report the basic data on the
stars 
used as background sources, in Table \ref{basic}. 
The spectral resolution is about 43000, 
i.e. the velocity resolution is about 
7 km s$^{-1}$ at the NaI D2 line wavelength.
\par
We used the very same spectra used for stellar abundance
analysis  in the above quoted papers and
we refer the reader to those papers for
details on the data reduction.
For our purpose the main problem  in our data is 
that the terrestrial Na I D emission lines, when
present, cannot 
be satisfactorily subtracted. This implies 
that the   lowest speed
local IS components are affected, sometimes severely. 
However, the components    in the Sgr dSph velocity range 
are always clean from the atmospheric emission.
The heliocentric corrections were always applied before 
summing the spectra.
\par\noindent
Another serious problem is posed by  the stellar absorption lines,
though  these may be reliably identified with the aid of a
synthetic spectrum, since atmospheric
parameters and chemical abundances of the
background sources are known. For each star we computed
a synthetic spectrum using the Linux version of
SYNTHE \citep{ATLAS,kurucz} and the same atmospheric
parameters and abundances given in \cite{B00}, \cite{Bonifacio}
or \cite{sbordone}, as appropriate. An example of synthetic  spectra
is shown in Fig. \ref{Fig2}, with this aid we rejected
possible identifications of interstellar lines for which
one of the components appeared to be 
severely blended with a 
stellar absorption line.  
The wavelength, column density and $b$ coefficient were therefore 
derived only when the lines were detected  in both components and 
beyond a 3$\sigma$ level over the synthetic spectra, i.e. only when 
the latter were not able to completely reproduce the features. 
The analysis was performed by means of the code 
{\tt fitvoigtminuit} 
(E. Caffau, 2004 private communication). 
This code, starting from initial guesses provided by the user, 
performs a voigt profile  
($\chi^2$) fit; prior to the computation
of the ($\chi^2$) the computed profile is convolved
with an instrumental profile, in our case
we used a gaussian profile with a FWHM
of 7 \kms, which is appropriate for our
data, as can be verified from
the profiles of unblended Th I lines in 
the calibration arc spectrum.  The minimum is found by the 
{\tt MINUIT} routine \citep{james}. 
The fitting parameters for each component are four:
central wavelength, column density, 
$b$ factor and $gf$ value. One of the 
useful features of {\tt MINUIT}
is that any of the parameters can be held fixed, even all
four, in which case there is no fitting, the routine
simply computes a line for the input parameters.
In our case the $gf$ values, which are well known
for the Na I D lines, were of course always kept fixed
(see Table \ref{atomic}).
\begin{table}
   \caption{atomic parameters used in the fitting procedure} 
\label{atomic}
\centering
\begin{tabular}{lcc}
\hline
line   & $\lambda(\AA)$ & f value \\
NaI D1 &  5889.9510 & 0.65459 \\
NaI D2 &  5895.9240 & 0.32732 \\
\hline
\end{tabular} 
\end{table}

At this relatively low resolution
there is no need to take into account the hyperfine
structure of the Na I D lines, which simply results
in a slightly larger $b$ value.
In the cases in which the detected IS lines are contaminated
by the stellar absorptions (clear examples are the
high velocity components shown in Fig.\ref{Fig2}), 
to properly recover all the parameters of the
IS lines (central wavelength, $b$ and column density)
it would
be necessary to fit {\em simultaneously} the stellar and interstellar
lines. Although this is possible, in principle, using the
stellar synthetic spectra,  we felt that considering the
rather low S/N ratio in the spectra and the additional uncertainty
introduced by the modelling of the stellar absorptions the results
would be affected by a  large error, too large in fact to be
a useful measure. 
Instead, in the exploratory spirit of the present work,
we decided in these cases to {\em assume} the values
of $b$ and column density and fit only the
central wavelength of the IS absorption. 
Although this procedure is somewhat arbitrary
it is fairly robust.  As long
as column density and $b$ are fixed arbitrarily, but
in   such a way
that the resulting synthetic spectrum is not 
obviously incompatible with the observed spectrum,
the fitted centroid of  the component 
does not change  by more than 1 \kms.
This is clearly sufficient for the exploratory
purpose of the present study. 
%The only concern could be the wavelength 
%range in which we are computing the synthetic spectrum, i.e. the
%stellar absorptions must be excluded very carefully in order to avoid
%contaminations.
The results of our analysis are 
reported in Table~\ref{tabstars}.
When $b$ and column density were
kept fixed,  the symbol 
* appears near the  log N values. 
\par\noindent
A few comments are in order here. The available
resolution is
somewhat limited for a detailed analysis of the
kinematic structure of the ISM. It is quite likely
that in higher resolution  observations
the ``components'' we fitted split into several
narrower components. The number of components to
fit to a given feature is somewhat arbitrary
and, for the above reason, may not have particular
physical meaning. We used the criterion to use
the minimum number of components in the fit.
It is thus clear that 
we do not give particular physical
significance to the column densities and
$b$ factors obtained,  keeping in mind
that they reflect the combined effect of 
several gas clouds.
In particular $b$ factors are not very significant;
most of the values we find are well below our
instrumental resolution, and thus, not very well
constrained. We stress here that
the FWHM of all the detected features is {\em above}
the instrumental resolution. A very low
fitted value of $b$ simply reflects the fact
that the FWHM of the feature is only very
slightly larger than the instrumental
profile, however at the S/N ratios of the present
data all the values of $b$ up to  7 \kms ~
provide almost equally acceptable fits.
What {\em is} significant is that
we do not observe $b$ factors in large excess of
our instrumental resolution. This allows to conclude
that there is no strong turbulence within the clouds
and that the velocity dispersion of groups
of clouds which cluster in velocity space
(our ``components'') is less than our
instrumental resolution, i.e. 7 \kms.

\section{Results and discussions}

From Table~\ref{tabstars} we can infer 
the presence of two distinct groups of IS components: 
one which we call  ``local'', ranging 
from -43 \kms ~ to 50 \kms, and a high velocity group ranging 
from 150 \kms ~ to 165 \kms. 
   \begin{figure*}[ht]
   \centering
   \includegraphics[width=14cm]{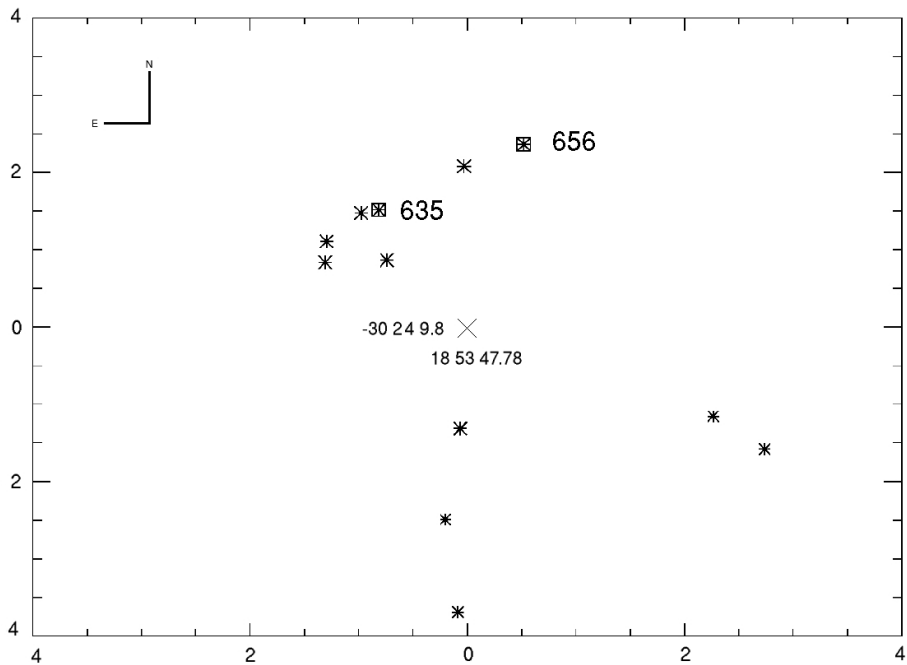}
      \caption{Map of the observed stars in Sgr dSph, the stars in squares are
      those in which high velocity absorptions have been detected, 
      the cross ($\times$) indicates the center of the field,
      labeled by its coordinates.       Scales are in
      arcminutes.
              }
         \label{Fig1}
   \end{figure*}

   \begin{figure*}
   \centering
   \includegraphics[width=16cm]{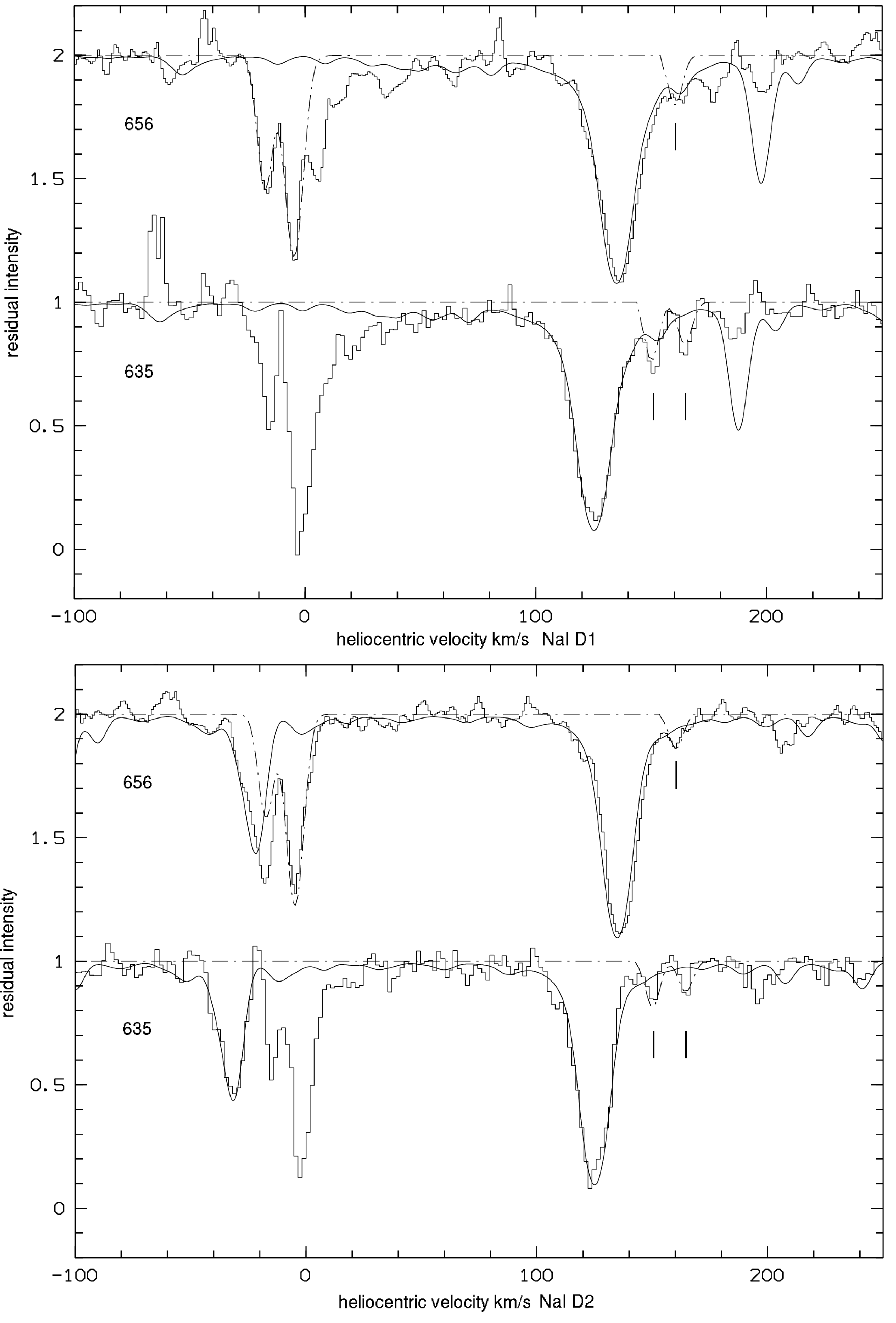}
      \caption{Spectra of stars showing high velocity components. 
The dashed-dotted line
      is the synthetic IS fit 
whereas the continuous line is the synthetic stellar spectrum.
      The local
      components of \# 635 are severely contaminated by 
the sky and no fit was reliable.
              }
         \label{Fig2}
   \end{figure*}

\subsection{The local components}
In most cases the sky 
contaminations prevent us from precise column densities or $b$ measures, 
nevertheless the velocity determinations are reliable. 
\par\noindent
We underline the 
presence of the $\simeq$-43 km/s and $\simeq$0 
\kms ~ towards all the Terzan7 stars,
which are probably due to a 
local cloud covering the small angular spread of these lines of sight. 
The column densities and $b$ values are consistent with this scenario. 
\par\noindent
The component around --15 \kms ~  is 
also quite ubiquitous, we find it towards 7 out of 12 lines of sight towards 
the Sgr dSph. 
The \citet{Redfield} model for the G and
Local Interstellar Cloud in these directions, 
gives velocities of --24.27 \kms ~and --22.2 \kms ~
respectively,
at our resolution these two components would appear
as a unique feature at $\sim -23$ \kms.
It is not clear whether the component around 
$- 15$ \kms found by us could be indeed compatible
with the  \citet{Redfield} model. As
mentioned above, 
we had a heavy contamination by 
atmospheric NaI emission lines, and considering that the
heliocentric corrections are around --8,--10 \kms, 
a possible systematic error in our velocity measure,
due to an incorrect subtraction of the atmospheric
emission,
leaves open the question of the 
identification with the above components predicted by the
\citet{Redfield} model.
\par\noindent
Another component is revealed between --7 and --2 \kms. This 
velocity spread is not very significant, 
since our velocity resolution is $\simeq$7 \kms, perhaps it could 
be related to  the 0 \kms ~component toward Terzan 7.
The column densities and $b$ values (where reliable) 
are not as homogeneous as  those observed
in the direction of Terzan 7.
\par 
Other components behave in the 
same ubiquitous manner, 
temptatively we can identify the following ones:
9-15 \kms, 18-28 \kms, 34-36 \kms and
40-45 \kms the latter two component are present towards 3 stars.
Other absorptions are
observed  towards less than 3 lines of sight, 
their distances are probably greater then the previous ones, 
but we cannot infer any more about them. 
The evident absorption features of star \#635
have not been fitted due to severe sky 
contaminations of both D1 and D2 lines. 

\balance
\subsection{The high velocity components}
As already said, this is the first investigation towards 
the Sgr dSph, therefore the finding of high velocity 
IS absorptions could reveal the presence of gas in it, or in 
the Galactic Halo on the other side of 
Galactic centre. 

Only three components of velocities 150, 160, and 165 \kms ~have been
found towards only two lines of sight, as reported in Fig.~\ref{Fig1}. 
In particular, the 150 and 165 \kms ~ are revealed towards  star \# 635 
as shown in Fig.~\ref{Fig2}.
The non ubiquitous nature of these 
absorptions should indicate the great distance of the 
absorbing gas, given the relative angular 
clumpiness of our sample, and in particular of 
these two stars. Nearby 
clouds should cover more than 2 or 3 lines of sight, but, 
at the other side of our Galaxy, their angular sizes could not.
In particular, the nearest star to \# 635 is 
\# 927, which shows no detectable
IS absorptions at these velocities, its angular separation
corresponds to 
a linear separation of 
$\simeq$2.9pc at the distance of Sgr dSph. The non detection
could be explained by three causes: 
\begin{enumerate}
\item since the radial velocity of star 
\# 927 is 18.8 km/s higher than that of the \# 635, 
it is obvious from Fig.\ref{Fig2} that 
the stellar features could mask completely 
the IS ones in this velocity range.
\item 
since from the analysis of 
\citet{Bonifacio} the \# 927 is more metal rich than \# 635 by a factor of two,
it shows stronger stellar absorptions 
just around these components and this probably could, by itself, 
prevent us from detecting them even if they were present;
\item 
the clouds are very clumpy, of size less than 2.9pc.   
\end{enumerate}
Arguments  1) and 3) apply also to star \# 628, which falls 
exactly between the \# 635 and \# 656, and has a larger
angular distance than  the angular distance
of star \# 927  from 
star \# 635, from either star  \# 635  or star \# 656.
The lack of HI detections in the Sgr dSph 
direction, \citep{Burton}, could 
be well explained by the small angular size of case 3). 
Moreover, it is very interesting to note that these 
velocities are always greater than the stellar  ones, 
implying that the gas is ``falling'' towards them.

\section{Conclusions}
The presence of high velocity absorptions 
towards 2 out of a sample of 12 stars of the Sgr dSph is the main 
result of our analysis. 
Given the \citet{Burton} upper limit on HI mass over the area, 
one can derive an upper limit for the HI column density of 
$\log N(HI)\leq18.87$. 
We stress the fact that the $21'$ 
radio beam integrates the signal over
an area which is considerably 
larger  than our field.  The Sgr stars
were initially found as radial velocity
members on an EMMI MOS frame
\citep{B99} which had a $9'$ side. 
In spite of this it is interesting to derive
an estimate of an upper limit on 
the corresponding NaI column density.
Such an estimate is clearly quite uncertain,
in fact  one
has to assume a Na/H ratio, an ionization 
balance and a dust depletion factor. Rather than relying on
theoretical estimates, we can assume the 
physical state of the gas to be similar to what
observed by \citet{Molaro} 
towards the Magellanic Clouds. They  found a value of $\log(NaI/HI)=-7.79$
for Intermediate Velocity Clouds,
thus we could expect  $\log NaI\leq 11.08$.
The fact that we assume higher column densities, but still compatible
with the observations, 
suggests one out of three possibilities:
either our assumed
column densities are too high by 
$\sim 0.5-0.8$ dex,
or  our detected high velocity absorptions arise in a clumpy
medium which prevents detection of HI emission
with a radio beam of $21'$,
or the physical conditions in the
gas are different from what assumed to derive the
upper limit on NaI column density and lead to $ \log(NaI/HI)>-7.0$.
Since this galaxy reveals a lack of gas, 
probably stripped away during the last
perigalactic passage, it is also possible that the interaction 
with the Milky Way caused some gas to
fall onto Sgr \citep[see][]{Putman}. 
Although our  conclusion has to be
considered preliminary, since 
it is based on only two
background stars and we have no indication
on the actual distance of the clouds, 
it is interesting to 
consider the implications of this finding,
assuming that the gas is at the same distance as Sgr.
In the first place it shows that there is 
a small amount of gas in Sgr, possibly clumpy and
consisting of many small clouds. This gas may be 
the residual of the  gas which was used up
in the last star formation episode of Sgr,
which gave birth to the young metal-rich population 
found by \citet{Bonifacio},
the rest being now in the form of stars or lost
along the orbit.  
In the second
place since this gas is chemically enriched
and ``falling'' towards Sgr it could
well be gas which has been stripped
from the Milky Way. If at each passage
Sgr is actually capable of
``refueling'' with metal enriched gas
stripped from the Galactic disc, this
could be a means to maintain  
the star formation episodes over a very long
period of time, as testified by the
very large metallicity spread of Sgr
($\rm -3\le[Fe/H]\le 0.1$ \citealt{zaggia}),
even without a large reservoir of local gas.
Moreover it could explain the anomalously
high metallicity of Sgr, compared to its 
luminosity; perhaps Sgr is not so chemically
evolved because it retains efficiently 
the ejecta of its own SNe, but simply 
because it ``steals'' enriched material
from the Milky Way.
These considerations are still
highly speculative, however
they provide an important thrust
for the
continuation of such an investigation. 
Data for more lines of sight using
hotter background sources which span 
a range of distances from the Galactic
disc to Sgr are badly needed to test these hypothesis.

\begin{acknowledgements}
We are grateful to E. Caffau for providing
the {\tt fitvoigtminuit} code and to 
S. Zaggia for providing accurate coordinates of the Terzan 7 stars.
We are also in debt with the referee Dr. J. Linsky because the article has 
been improved thanks to his constructive comments.
This research was done with support from the
Italian MIUR COFIN2002 grant
``Stellar populations in the Local Group
as a tool to understand galaxy formation and evolution'' (P.I. M. Tosi).
\end{acknowledgements}


\begin{thebibliography}{}
\bibitem[Bonifacio et al. (1999)]{B99}
Bonifacio P., Pasquini L., Molaro P., Marconi G., 1999, \apss , 265, 541
\bibitem[Bonifacio et al. (2000)]{B00}
Bonifacio P., Hill V., Molaro P., Pasquini L., Di Marcantonio P.,
Santin P.  2000, \aap , 359, 663
\bibitem[Bonifacio et al. (2004)]{Bonifacio} Bonifacio P. et al. A\&A, 414, 503, 2004
\bibitem[{Buonanno et al. (1995)}]{Buonanno}
Buonanno R. et al.  1995, AJ 109, 663
\bibitem[Burton and Lockman (1999)]{Burton} Burton W and Lockman F.J. A\&A, 349, 7, 1999
\bibitem[James (2001)]{james} James, F., 1998 MINUIT, Reference Manual, Version 94.1, CERN, Geneva, Switzerland
\bibitem[Koribalski et al. (1994)]{Koribalski} 
Koribalski B., Johnston S. and Otrupeck R., MNRAS, 270, 43 1994
\bibitem[{Kurucz (1993)}]{kurucz}
Kurucz, R. L. 1993, CD-ROM 13, 18~
\href{http://kurucz.harvard.edu}{http://kurucz.harvard.edu}
\bibitem[Marconi et al.(1998)]{Marconi} Marconi, G., Buonanno, 
R., Castellani, M., Iannicola, G., Molaro, P., Pasquini, L., \& Pulone, L.\ 
1998, \aap, 330, 453 
\bibitem[Molaro et al.(1993)]{Molaro} Molaro, P. et al. A\&A, 274, 505, 1993
\bibitem[Putman et al. (2004)]{Putman} Putman et al. Ap.J., 603, L77, 2004 
\bibitem[Redfield and Linsky (2000)]{Redfield} Redfield S. \& Linsky J.L. Ap.J., 534, 825, 2000
\bibitem[Sbordone et al. (2004)]{sbordone}Sbordone, L.,
Bonifacio, P., Marconi, G., \& Buonanno, R.\ 2004, MSAIt, 75, 396
\bibitem[Sbordone et al. (2004)]{ATLAS}Sbordone, L.,
Bonifacio, P., Castelli, F., \& Kurucz, R.L.\ 2004, MSAIS, 5, 93 
\bibitem[Zaggia et al. (2004)]{zaggia} Zaggia et al. MSAIS, 5, 291
\end{thebibliography}
\end{document}